\documentclass{elsart}
\usepackage{times}
\usepackage{epsfig}
\begin{document}
\bibliographystyle{roman}

\journal{Astroparticle Physics}

%%%%%%%%%%%%%%%%%%%%%%%%%%%%%%%%%%%%%%%%%%%%%%%%%%%%%%%%%%%%%%%%%%%%%%%%%%%%%%%
\def\hb{\hfill\break}
\def\MeV{\rm MeV}
\def\GeV{\rm GeV}
\def\TeV{\rm TeV}

\def\m{\rm m}
\def\cm{\rm cm}
\def\mm{\rm mm}
\def\lam{$\lambda_{\rm int}$}
\def\rad{$X_0$}
 
\def\NIM{Nucl. Instr. and Meth.~}
\def\etal{{\it et al.~}}
\def\eg{{\it e.g.,~}}
\def\ie{{\it i.e.~}}
\def\cf{{\it cf.~}}
\def\etc{{\it etc.~}}
\def\vs{{\it vs.~}}
%%%%%%%%%%%%%%%%%%%%%%%%%%%%%%%%%%%%%%%%%%%%%%%%%%%%%%%%%%%%%%%%%%%%%%%%%%%%%%%
\begin{frontmatter}
\title{PeV Cosmic Rays: A Window on the Leptonic Era?}

\author{Richard Wigmans\thanksref{Corres}}

\address{Department of Physics, Texas Tech University, Lubbock, TX 79409-1051, USA}
\thanks[Corres]{Email wigmans@ttu.edu, fax (806) 742-1182.}

%%%%%%%%%%%%%%%%%%%%%%%%%%%%%%%%%%%%%%%%%%%%%%%%%%%%%%%%%%%%%%%%%%%%%%%%%%%%%%%
\begin{abstract}
It is shown that a variety of characteristic features of the high-energy hadronic cosmic ray spectra, in
particular the abrupt changes in the spectral index that occur around 3 PeV and 300
PeV, as well as the corresponding changes in elemental composition that are evident from kinks in the
$\langle X_{\max}\rangle$ distribution, can be explained in great detail 
from interactions with relic Big Bang antineutrinos, provided that the latter have a rest mass of $\sim 0.5$
eV/$c^2$.

\end{abstract}

\end{frontmatter}

%%%%%%%%%%%%%%%%%%%%%%%%%%%%%%%%%%%%%%%%%%%%%%%%%%%%%%%%%%%%%%%%%%%%%%%%%%%%%%%
\section{Introduction}

The energy region between 1 and 10 PeV is an area of intense study in cosmic ray research.  
The all-particle cosmic-ray energy spectrum falls extremely steeply with energy.
In general, it is well described by a power law
\begin{equation}
{dN\over dE}~\sim~ E^{-n}
\label{cosray}
\end{equation}
with $n \approx 2.7$ for energies below 1 PeV.
%In order to display characteristic features of this spectrum which would
%otherwise be hard to discern, the differential energy spectrum has been
%multiplied by $E^{2.7}$ in this figure. 
The steepening that occurs between 1 PeV and 10 PeV, where the
spectral index $n$ changes abruptly from 2.7 to 3.0, is known as the {\em knee} of the cosmic ray spectrum. 

This phenomenon is generally  believed to contain key information about the origin of the cosmic
rays and about the acceleration mechanisms that play a role. Especially models in which the cosmic rays are
resulting from particle acceleration in the shock waves produced in Supernova explosions
have received much attention in the literature. Such models predict a maximum energy, 
proportional to the nuclear charge $Z$ of the particles \cite{Blanford}. In the context of these models, the
knee is assumed to be associated with this ($Z$-dependent) maximum and the corresponding cutoff phenomena. In
the past years, major efforts have been mounted to determine the elemental composition of the cosmic
rays in the knee region. These efforts have revealed that the knee coincides with an abrupt change in the elemental composition
of the cosmic rays. 

We would like to point out that the high-energy cosmic ray spectra contain several other remarkable features.
For example, there
is a significant second knee in the energy spectrum at $\sim 300$ PeV, which coincides with an abrupt change in the elemental
composition as well. Even though these features are experimentally well established, they have received little or no attention in
the literature, presumably because they do not fit in the context of the aforementioned shock wave acceleration models.  

In this paper, we show that {\em all} measured features of the cosmic ray spectra in the energy range from
$10^{14}$ eV to $10^{18}$ eV are in detailed 
%and spectacular 
agreement with the predictions of a completely
different model. In this model, interactions between the cosmic rays and $\bar{\nu}_e$ remnants from the Big
Bang play a crucial role. If this is correct, then the experimental cosmic ray data make it directly possible to determine the
rest mass of these neutrinos. The result, $m_{\nu_e} = 0.5 \pm 0.2$ eV/$c^2$, falls inside the narrowing window still allowed by
explicit measurements of this important parameter.
If the role of interactions with relic neutrinos is indeed as important as suggested by the cosmic ray data,
then this also provides crucial information about the possible origin of the PeV cosmic rays and about the
acceleration mechanisms.
  
This paper is organized as follows. In Section 2, we review the key elements of the experimental
cosmic ray information in the energy range from $10^{14}$ eV to $10^{18}$ eV. In Section 3, we describe how
interactions with relic neutrinos might explain these phenomena. In Section 4, a possible scenario for the
origin of PeV cosmic rays is discussed. Conclusions are given in Section 5.

\section{Cosmic rays in the 0.1 -- 1000 PeV range}

All measurements in this energy range have been performed in extensive air-shower experiments.
The detectors measure the \v{C}erenkov light, the scintillation light and/or the charge produced by the
shower particles generated in the absorption process that takes place in the Earth's atmosphere. Some
experiments, \eg the Fly's Eye \cite{Cassiday}, are capable of reconstructing the shower profile in the
atmosphere. This may provide important information about the type of particle that initiated the shower.

\subsection{The knee}
\begin{figure}[htb]
\epsfysize=9cm
\centerline{\epsffile{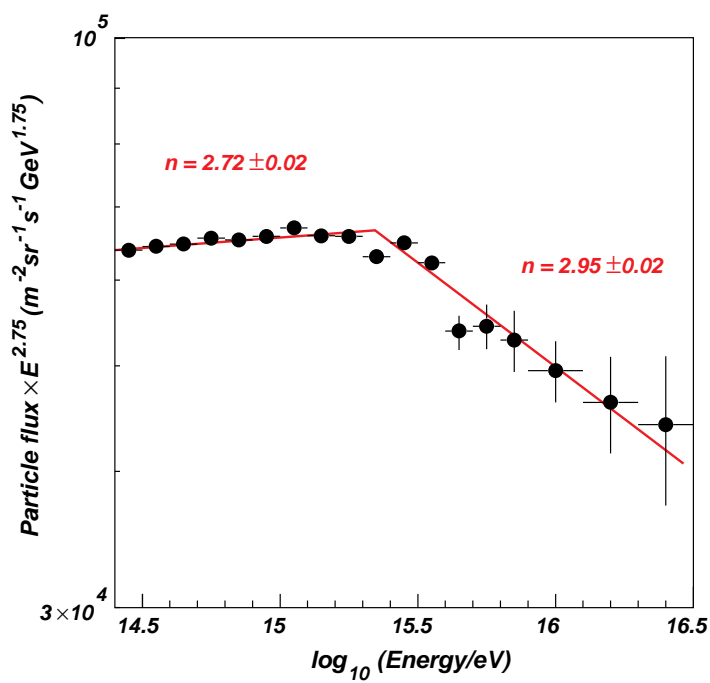}}
\caption{\small
The all-particle energy spectrum of cosmic rays, measured by the CASA-BLANCA experiment
\protect\cite{blanca}.}
\label{knee1}
\end{figure}

\vskip-5mm
In the past 20 years, about a half-dozen experiments have measured the cosmic ray spectra in the energy range 
from 1 -- 10 PeV. The existence of a kink in this area has been very well established. The different experiments
agree on the fact that the observed change in the spectral index $n$ is very significant and occurs very
abruptly. 
As an example, Figure \ref{knee1} shows the data from the CASA-BLANCA experiment \cite{blanca}, which
recently performed measurements in the energy range from 0.3 PeV to 30 PeV at the Dugway site in Utah (U.S.A.),
near the location of the Fly's Eye detectors \cite{hires}. The spectral index was found to change from $n = 2.72 \pm 0.02$
at energies below 2 PeV to $n = 2.95 \pm 0.02$ above 2 PeV. Similarly significant kinks were reported by other experiments, 
\eg Akeno \cite{akeno}, Tibet AS$\gamma$ \cite{tibet} and DICE \cite{dice}.

The abruptness of the change in the spectral index clearly suggests that some kind of threshold is crossed. However, the precise
value of the threshold energy varies from one experiment to the next. This is undoubtedly a consequence of differences in the
energy calibration methods of the experimental equipment that were applied in the different experiments. Given the absence of
a reliable calibration source with precisely known energy in the PeV regime, it is no surprise that the absolute energy
scales differ by as much as a factor of two. The reported values for the threshold energy range from 2 PeV (\eg for
CASA-BLANCA) to 4 PeV (Akeno). In the following, we will adopt a value of 3$\pm$ 1 PeV. 
          
\subsection{The kink near 300 PeV}
\vskip-5mm
Several extensive air-shower experiments that have
studied the cosmic ray spectrum at the highest energies have reported a kink in the
area around $\log{E} = 17.5$. The Fly's Eye experiment, which obtained the largest event
statistics, observed a change in the spectral index from $3.01 \pm 0.06$ for energies
$< 10^{17.5}$ eV to $3.27 \pm 0.02$ for energies $10^{17.5} < E < 10^{18.5}$ eV \cite{fleye}.
The Haverah Park experiment also reported a kink at $\log{E} = 17.6$, with the spectral index
changing from $3.01 \pm 0.02$ to $3.24 \pm 0.07$ \cite{haverah}. 
\begin{figure}[htb]
\epsfysize=8cm
\centerline{\epsffile{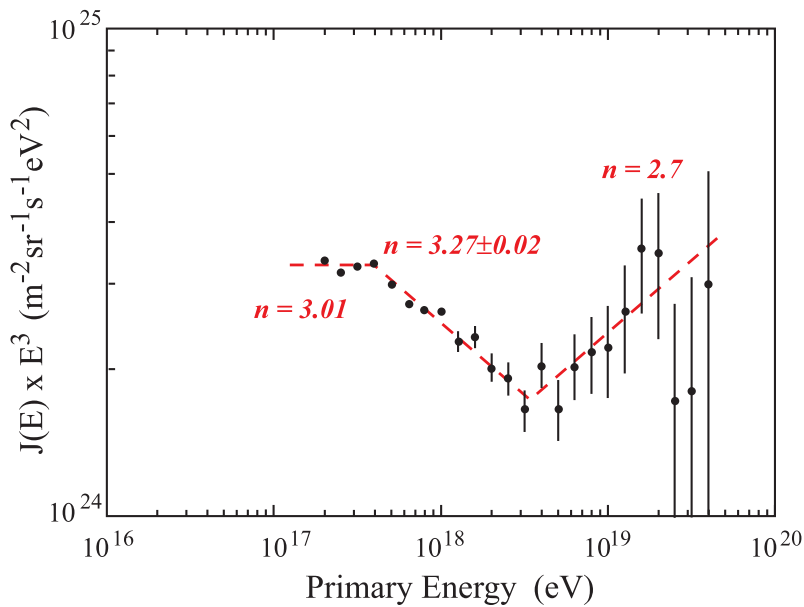}}
\caption{\small
The all-particle energy spectrum of cosmic rays, measured by the Fly's Eye experiment
\protect\cite{fleye}.}
\label{fly}
\end{figure}

The Fly's Eye results are shown in Figure \ref{fly}. In order to better discern the characteristic features,
the differential energy spectrum has been multiplied by $E^3$, as opposed to $E^{2.75}$ in Figure \ref{knee1}. 
Figure \ref{fly} shows several other interesting features, such as the {\em ankle} near
$\log{E} = 18.5$, as well as several events above the GZK limit \cite{GZK}. The authors of Reference \cite{fleye} concentrated
their attention entirely on these phenomena near the high-energy end of their experimental reach. 
However, the change of the spectral index near their low-energy limit is also very interesting. 

The fact that the Fly's Eye detectors lacked sensitivity below 100 PeV limited the significance of
their measurement of the change in the spectral index to about 4 standard deviations.  However, the fact that several other
experiments have measured $n$ values of 2.95 - 3.00 with a precision of the order of 0.02 in the energy range from 5 -
100 PeV \cite{blanca,haverah}, while Fly's Eye measured $n = 3.27 \pm 0.02$ for energies between 300 and 3000 PeV makes the
overall significance of this kink comparable with that of the knee around 3 PeV.

\subsection{The elemental composition}
\vskip-5mm
Since the detectors in extensive air-shower experiments are located behind an absorber with a thickness of about 10 nuclear
interaction lengths (the Earth's atmosphere), it is usually impossible to determine the identity of the incoming cosmic particle 
event by event. However, it is in some experiments possible to distinguish protons, $\alpha$ particles and heavier
nuclei on a statistical basis. In experiments such as Fly's Eye \cite{fleye}, this is done by measuring the shower profile in the
atmosphere. These profiles are, on average, quite different for the mentioned constituents of the cosmic ray spectra.
In other experiments, \eg KASCADE \cite{kascade}, a large number of different shower characteristics are used simultaneously in
a neural network that is trained to assign probabilities that the event was initiated by a proton, an $\alpha$ particle or a
heavier nucleus on the basis of the experimental information.

One experimental parameter that is frequently used in this context is the average depth in the atmosphere at which the
shower reaches its maximum intensity, $\langle X_{\rm max} \rangle$. At a given energy, $\langle X_{\rm max} \rangle$ is
larger for protons than for heavier ions, and its value decreases as the nuclear charge $Z$ of the projectiles increases. There
are two reasons for these effects:
\begin{enumerate}
\item The {\em nuclear interaction length} ($\lambda_{\rm int}$), \ie the average distance the primary particle penetrates into
the atmosphere before undergoing a nuclear interaction, is proportional to $A^{-2/3}$. Therefore, protons penetrate, on average,
much deeper into the atmosphere than do heavier nuclei.
\vskip 2mm
\item The {\em particle multiplicity} is smaller in reactions initiated by protons than in those initiated by heavier
nuclei. Therefore, the energy of the incoming proton is transferred to a smaller number of secondaries, which carry
thus, on average, more energy than the secondaries produced in reactions initiated by heavier ions of the same primary energy.
And since the depth of the shower maximum increases with energy, the showers developed by the secondaries
in proton-induced reactions reach their maximum intensity farther away from the primary vertex than in case of showers
induced by heavier ions.        
\end{enumerate}

In summary, showers induced by protons of a given energy start later and peak at a larger distance from the primary vertex
than showers induced by heavier ions of the same energy. These effects can be quantitatively estimated on the basis of the
well known characteristics of showers at lower energy, for example in the following way  \cite{book}.

The Particle Data Group lists the nuclear interaction length for protons in air as 90 g cm$^{-2}$ \cite{PDG}. Combined with
the $A^{-2/3}$ cross section dependence, this leads to estimates of $\lambda_{\rm int} = 36$ g cm$^{-2}$ for $\alpha$s and
$\lambda_{\rm int} = 6$ g cm$^{-2}$ for iron nuclei in air. Therefore, effect 1 listed above will cause $\langle X_{\rm max}
\rangle$ for proton-induced showers to be 54 g cm$^{-2}$ larger than for $\alpha$-induced showers and 84 g cm$^{-2}$ larger than
for showers induced by iron nuclei.

\begin{figure}[htb]
\epsfysize=8.38cm
\centerline{\epsffile{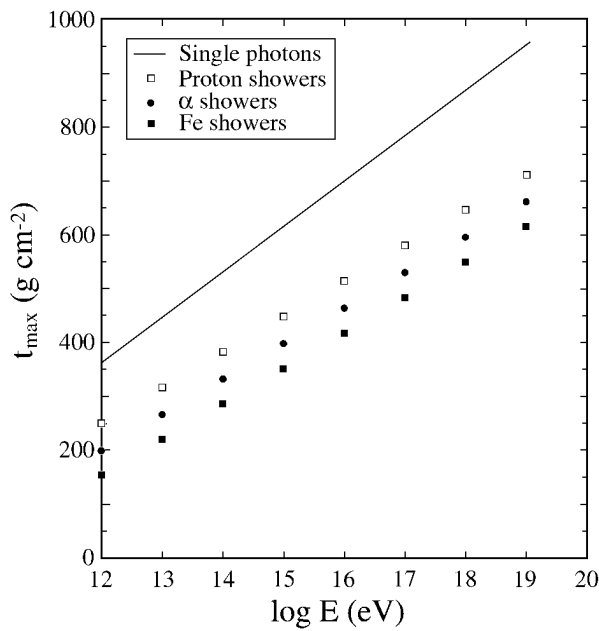}}
\caption{\small
The average depth of the shower maximum, calculated from the starting point of the showers, as a function of energy. Shown are
the results for showers induced by single photons (Equation \ref{showermax}) and for the electromagnetic component of
showers induced by protons, $\alpha$s and Fe nuclei in the atmosphere. The latter were calculated on the basis of
multiplicity assumptions discussed in the text.}
\label{emair}
\end{figure}

To estimate the second effect, it is important to realize that the maximum intensity in air
showers is reached, in first approximation, at a depth where the electromagnetic showers developed by photons from decaying
$\pi^0$s produced in the {\em first} nuclear reaction reach their peak intensity. This is a consequence of the fact that the
interaction length for charged pions in air is only $\sim 4$ times larger than the radiation length. In dense absorber
materials, these two quantities may differ by as much as a factor of 30 and, as a result, hadron showers in dense detectors,
such as calorimeters used in particle physics experiments, have very different characteristics \cite{book}.      
The maximum intensity in a shower induced by a photon is approximetely reached at a depth
\cite{longo}:       
\begin{equation}
t_{\rm max} = \ln y + 0.5
\label{showermax}
\end{equation}
where $t_{\rm max}$ is expressed in radiation lengths ($X_0$) and $y$ is the photon energy,
expressed in units of the critical energy ($\epsilon_c$). This relationship is graphically represented by the solid line in
Figure \ref{emair} for photon-induced showers in air ($X_0 = 36.7$ g cm$^{-2}$, $\epsilon_c = 87$ MeV).  

In order to calculate the shower maximum in hadron-induced showers, we have to know the average fraction of the energy
of the incoming particle that is transferred to individual photons. This may be derived from the multiplicity
distributions measured in accelerator-based experiments. 
\begin{figure}[htb]
\epsfysize=7cm
\centerline{\epsffile{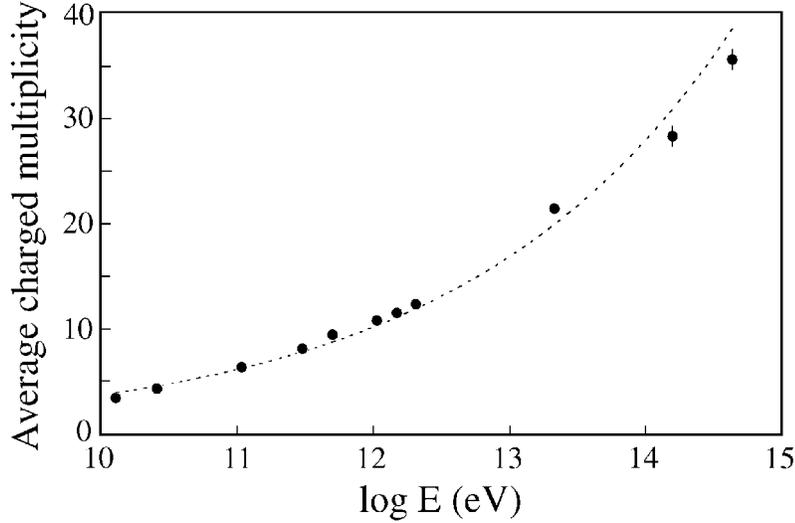}}
\caption{\small
The average charged particle multiplicity measured in $pp$ reactions, as a function of energy.
The dashed line represents the fit used to extrapolate these data to higher energies. See text for details.}
\label{ppmulti}
\end{figure}

Figure \ref{ppmulti} shows the average multiplicity of charged particles produced in $pp$ interactions as a function of
energy \cite{PDG,ppsig}. Most of these data come from collider experiments and we have Lorentz-transformed these data
to a fixed-target geometry. The dashed line is the result of an exponential fit and, for lack of a better method, we have used
this fit to extrapolate to higher energies. The average photon energy in the proton-induced showers was calculated as
follows. For example, at $\sqrt{s} = 546$ GeV, which corresponds to a fixed-target energy of $1.59 \cdot 10^{14}$ eV,
the measured charged multiplicity was, on average, $28.3 \pm 1.0$ \cite{ppsig}. Assuming that equal numbers
of $\pi^+$s, $\pi^-$s and $\pi^0$s are produced in the nuclear interactions, the total multiplicity is thus 42.5 and since a
$\pi^0$ decays into 2 photons, these photons carry, on average, 1/85 of the energy of the incoming proton. The average distance
from the starting point of proton showers to the shower maximum was calculated at this energy on the basis of Equation
\ref{showermax}, using a photon energy of $1.59 \cdot 10^{14}/85~=~1.87 \cdot 10^{12}$ eV. The other proton points (the
open squares in Figure \ref{emair}) were found in a similar way. 

The points for showers induced by $\alpha$ particles and by iron nuclei were found by assuming that the other nucleons
in the projectile would start simultaneous showers in the first nuclear interaction and that the initial energy would thus be
shared among a correspondingly increased number of secondaries. This assumption is based on experimental observations in
high-energy heavy-ion scattering scattering experiments at CERN and Brookhaven.
In the case of $\alpha$-induced showers, we therefore
increased the multiplicity by a factor of 4 and in the case of iron nuclei by a factor of 14, since the target nucleus
(predominantly $^{14}$N) only contains 14 nucleons. This simplifying approach overestimates the multiplicities. Therefore,
the  differences between showers induced by the different nuclei shown in Figure \ref{emair} represent an upper
limit.

Figure \ref{emair} shows a logarithmic energy dependence of the shower maxima.  
This trend is somewhat modified when the effects of re-interacting charged pions are taken into account. Looking at Equation
\ref{showermax} and considering that $\lambda^\pi_{\rm int} \approx 4 X_0$
\footnote{The interaction length for pions is larger than that for protons in the same material. Differences of 20\% - 50\% have
been reported in the literature \protect\cite{book}}, 
we see that such effects will tend to shift the shower maximum to a larger depth if the average charged particle multiplicity is
less than $\sim 50$ ($e^4$). And since the average multiplicity increases with energy, this effect shifts the shower
maximum more for lower-energy cosmic rays than for the highest-energy ones. In calculating the effects of re-interacting pions
in the second, third, and higher generations of the shower development, we also have to take into account the fact that, as the
pions become less energetic, they are also more likely to decay rather than to re-interact. For example, a 100 GeV $\pi^+$
produced at a depth of 150 g cm$^{-2}$ has comparable probabilities to decay and to interact in the atmosphere. At higher
altitude, the decay probability increases, at lower altitude the particle is more likely to interact.
\begin{figure}[htb]
\epsfysize=8.5cm
\centerline{\epsffile{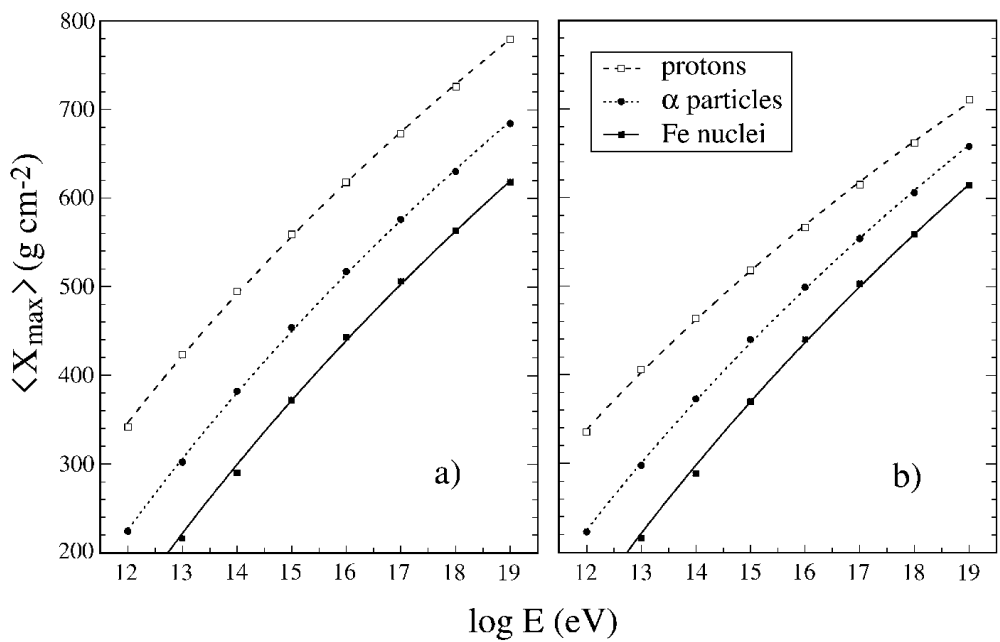}}
\caption{\small
The average depth inside the atmosphere at which the cosmic ray showers reach their maximum intensity. Results of calculations
that were performed for a constant cross section ($a$) and for an energy dependent cross section ($b$). See text for details.}
\label{hadairmax}
\end{figure}

Figure \ref{hadairmax}a shows the results of a hand-based calculation, in which we have taken these effects into account for 4
generations of particle multiplication. The average depth of the shower maximum increases slower than logarithmically with
energy. The curves for protons, $\alpha$-particles and iron nuclei run more or less parallel to each other. The latter
tendency changes when we also take into account the effect of a possible increase of the total cross section for the primary
nuclear interactions with energy. According to the Particle Data Group \cite{PDG,lbl},  
the total cross section for $pp$ collisions gradually increases from 40 mb at 1 TeV to 120 mb at
10$^{18}$ eV\footnote{It should be noted that this trend hinges on the merits of a {\em
single} experimental data point}. Therefore, the nuclear interaction length in air decreases by a factor of 3 over this energy
range, from 90 g cm$^{-2}$ at 1 TeV to 30 g cm$^{-2}$ in the EeV range. The interaction lengths of the heavier nuclei are
probably affected similarly. In Figure \ref{hadairmax}b, we have taken this effect into account as well. Obviously, it 
reduces the $Z$ dependence of $\langle X_{\rm max} \rangle$ as the energy increases. 

\begin{figure}[htb]
\epsfysize=9cm
\centerline{\epsffile{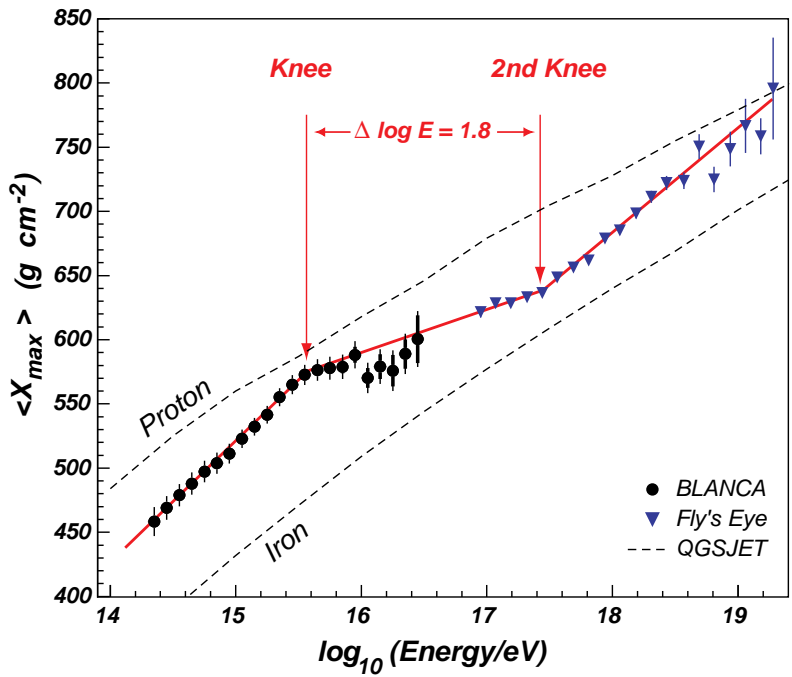}}
\caption{\small
The average depth of the shower maximum in the atmosphere as a function of the energy of the cosmic rays.
Experimental data were obtained with the detectors of the BLANCA \protect\cite{blanca} and Fly's Eye 
\protect\cite{fleye} experiments. The dashed lines represent model calculations for protons and iron nuclei
as the primary cosmic particles. See text for details.}
\label{xmax}
\end{figure}

The above discussion is intended as an introduction to the {\em experimental} $\langle X_{\rm max}\rangle$ data, which are shown
in Figure \ref{xmax}, together with the results of model calculations performed by the authors of the papers in which
these data were published \cite{blanca,fleye}. Its purpose is to demonstrate three things:
\begin{enumerate}
\item The conclusions drawn from simple considerations based on a fundamental understanding of shower development are confirmed
in detail by the results of very sophisticated and elaborate model calculations such as the QGSJET ones depicted in Figure
\ref{xmax}.
\vskip 2mm
\item The curve for $\alpha$-induced showers is located in between those for protons and iron in Figure
\ref{xmax}, somewhat closer to the iron curve than to the one for protons.
\vskip 2mm
\item Since all effects contributing to these model curves lead to smooth changes as a function of energy, the two kinks
observed in the experimental data (indicated by the arrows in Figure \ref{xmax}) represent a {\bf very remarkable phenomenon}.
\end{enumerate}

A kink in the $\langle X_{\rm max}\rangle$ distribution is strongly indicative for a threshold phenomenon, even more so than a
kink in the primary energy spectra. The cosmic rays consist of a mixture of protons, $\alpha$s and heavier nuclei. The
threshold concerns a process that {\em selectively} affects one of these components. Therefore, the elemental composition
abruptly starts to change when the threshold is passed.  
At low energies, protons are the most abundant particles. At the first
kink, protons drop selectively out of the mix and heavier species start to dominate. At the second kink, this process is
reversed: The protons start to come back and at the highest energies, they are again the most abundant components of the cosmic
ray spectrum.           
\begin{figure}[htb]
\epsfysize=8.38cm
\centerline{\epsffile{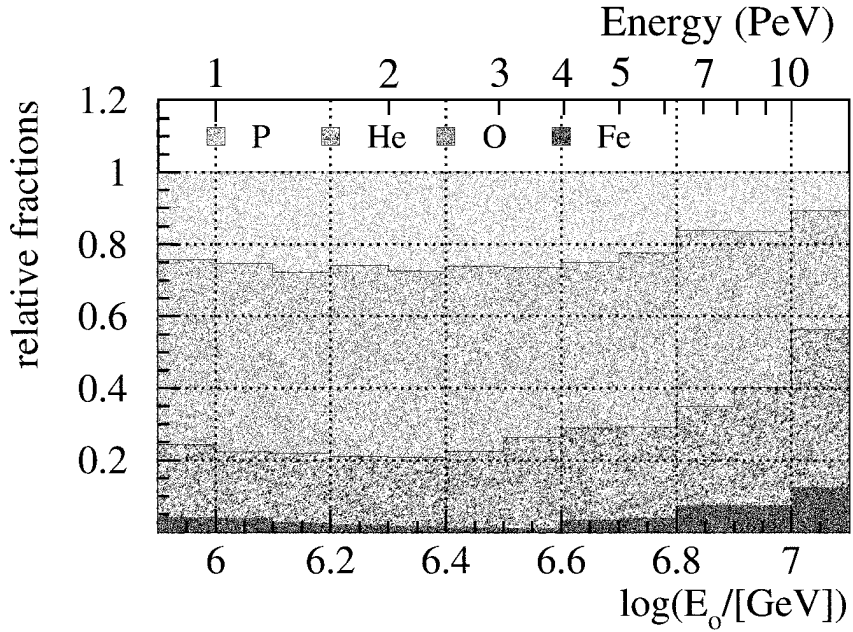}}
\caption{\small
The elemental composition of the cosmic ray spectrum in the region from $10^{15} - 10^{16}$ eV, measured 
by the KASCADE Collaboration \protect\cite{kascade}. The fractions of Helium, Oxygen and Iron have been normalized to the
Hydrogen content. See text for details.}
\label{kascade1}
\end{figure}

Additional evidence for this observation was provided by the KASCADE Collaboration \cite{kascade}. Figure \ref{kascade1} shows
the relative abundance of various elements as a function of energy, in the 1 -- 10 PeV range. The fractions of all elements have
been normalized to that of hydrogen. The figure shows an increase of the relative abundance of He, O and Fe beyond $\sim 4$
PeV. A {\em selective} reduction of the hydrogen content beyond the first kink would produce exactly this effect.  

It is remarkable that the two kinks in the $\langle X_{\rm max}\rangle$ distribution coincide with the two ``knees'' observed in
the all-particle energy spectrum itself (Figures \ref{knee1} and \ref{fly}). 
There is no {\em a priori} reason why that should be so. Note that there is {\em no} evidence for a kink in
the $\langle X_{\rm max}\rangle$ distribution near the ``ankle'' at $10^{18.5}$ eV. This ankle, where the spectral index $n$
changes from 3.3 back to its ``canonical'' value of 2.7 (see Figure \ref{fly}), is usually interpreted as the point where
cosmic rays of extragalactic origin start to dominate the galactic component. Below this energy, all charged particles are
confined by the galactic magnetic field and, therefore, the particles that are the subject of our study are  predominantly of
galactic origin.  

In Section 2.1, we ascribed differences in the precise energy at which the knee was found to different calibration procedures
applied by the experimental groups active in this field. It should be emphasized that the two experiments that produced the data
shown in Figure \ref{xmax} are located at the same site and that the two Collaborations have overlapping membership. Therefore,
it is unlikely that there are major systematic differences between the energy scales used in these two experiments. As a result,
the {\em energy gap} between the two kinks in Figure \ref{xmax} ($\Delta \log E = 1.8$) has most likely a much smaller systematic
uncertainty than the energies at which the individual kinks are located. As discussed in the next section, this energy gap plays
an important role in our explanation of all these experimental facts.

\section{Relic neutrinos}

\subsection{Properties}
\vskip-5mm
According to the Big Bang model of the evolving Universe, large numbers of (electron) neutrinos
and antineutrinos have been around since the beginning of time. In the very first second,
when the temperature of the Universe exceeded 1 MeV, the density was so large that the
(anti-)neutrinos were in thermal equilibrium with the other particles that made up the
primordial soup: photons, electrons, positrons and nucleons. Photon-photon interactions
created $e^+e^-$ pairs, which annihilated into photon pairs. Interactions between
(anti-)neutrinos and nucleons turned protons into neutrons and vice versa.

This {\em leptonic era} came to an end when the mean free path of the neutrinos become so large that their average lifetime
started to exceed the age of the Universe, $\sim 1$ second after the Big Bang.
Since that moment, the wavelengths of the (anti-) neutrinos have been expanding in
proportion to the size of the Universe. Their present spectrum is believed
to be a momentum-redshifted relativistic Fermi--Dirac distribution, and the number of particles per unit
volume in the momentum bin $(p,p+dp)$ is given by
\begin{equation}
N(p) dp~=~{{p^2 dp}\over{\pi^2 \hbar^3 [\exp (pc/kT_{\nu}) + 1]}}{({g_\nu\over 2})}
\label{nuspec}
\end{equation}
where $g_\nu$ denotes the number of neutrino helicity states \cite{TG}.
The distribution is
characterized by a temperature $T_\nu$, which is somewhat lower than that for the relic photons.
Since 
$(T_\nu/T_\gamma)^3 = 4/11$ and $T_\gamma = 2.726 \pm 0.005$ K \cite{COBE}, $T_{\nu}$ is expected to be 1.95 K.
The present density of these Big Bang relics is estimated at $\sim$ 100 cm$^{-3}$, for each neutrino flavor.
That is nine orders of magnitude larger than the density of baryons in the Universe. 

It is important to realize that, depending on their mass, these relic neutrinos might
be very {\em nonrelativistic} at the current temperature ($kT_\nu \sim 10^{-4}$ eV). Since they decoupled,
their momenta have been stretched by a factor $10^{10}$, from 1 MeV/$c$ to $10^{-4}$ eV/$c$. If their rest
mass were 1 eV/$c^2$, their maximum velocity would thus be $10^{-4} c$, or only 30 km/s.

The experimental upper limit on the mass of the electron antineutrino\footnote{In the following, we express masses in 
energy units, omitting the $c^{-2}$ factor.} was recently determined at 2.2
eV (95\% C.L.), from a study of the electron spectrum of $^3$H decay \cite{Mainz}. The
experimental results on atmospheric and solar neutrinos obtained by the Superkamiokande \cite{SuperK} and SNO \cite{SNO}
Collaborations suggest that neutrinos do have a nonzero rest mass. There is no experimental information that rules out a neutrino
rest mass in the bracket  0.1 -- 1 eV. 

Despite their enormous abundance, estimated at $\it{O} (10^{86})$ for the Universe as a whole, relic neutrinos have
until now escaped direct detection. The single most important reason for that is their
extremely small kinetic energy, which makes it difficult to find a process through which
they might reveal themselves. 

\subsection{How to detect relic neutrinos?}
\vskip-5mm
Let us imagine a target made of relic $\bar{\nu}_e$s and let us bombard this target with
protons.
Let us suppose that we can tune this imagined proton beam to arbitrarily high energies,
orders of magnitude beyond the highest energies
that can be reached in our laboratories.
Then, at some point, the proton energy will exceed the value at which the center-of-mass
energy of the $p-\bar{\nu}_e$ system exceeds the combined mass energy of a neutron and a positron.
Beyond that energy, the inverse $\beta$-decay reaction
\begin{equation}
p + \bar{\nu}_e \rightarrow n + e^+
\label{ibeta}
\end{equation}
is energetically possible.

The threshold proton energy for this process depends on the mass of the $\bar{\nu}_e$
target particles. If this mass is large compared to the $10^{-4}$ eV kinetic energy typically
carried by the target particles, this may be treated as a stationary-target problem, and
the center-of-mass energy of the $p-\bar{\nu}_e$ system can be written as
\begin{equation}
E_{\rm cm} = \sqrt{m_p^2 + m_\nu^2 + 2 E_p m_\nu} \approx \sqrt{m_p^2 + 2 E_p m_\nu} 
\label{com1}
\end{equation}
since $m_\nu \ll m_p$.  
When the experimental mass value of the proton (938.272 MeV) is substituted in
Equation \ref{com1} and the requirement is made that $E_{\rm cm} > m_n + m_e$ (940.077 MeV), this
leads to  
\begin{equation}
E_p m_\nu > 1.695 \cdot 10^{15}~ {\rm (eV)}^2 
\label{com2}
\end{equation}
This process will thus take place when
\begin{equation}
E_p ({\rm eV}) > {1.695 \cdot 10^{15}\over m_{\nu} ({\rm eV})} 
\label{com3}
\end{equation}

In our {\em Gedanken experiment}, this threshold would reveal itself through a decrease in the fraction
of beam protons that traversed the target without noticing its presence, as $E_p$ is
increased beyond the threshold. 
We notice that the knee at 3 PeV exhibits exactly the features that we expect to see in this
experiment: The particle flux suddenly starts to decrease when the threshold is passed.   
Therefore, we postulate the following hypothesis:

{\em The change of the spectral index in the all-particle cosmic ray spectrum at an energy of 
$\sim 3$ PeV is caused by the onset of the reaction $p + \bar{\nu}_e \rightarrow
n + e^+$, which becomes energetically possible at this point.}

This hypothesis necessarily implies (Equation \ref{com3}) that the mass of the electron neutrino equals 
$\sim 0.5$ eV. Also, the knee would have to be an {\em exclusive feature of the proton component} of the cosmic ray
spectrum, if the hypothesis were correct. Beyond 3 PeV, one would thus expect to see a gradual drop in, for
example, the $p/\alpha$ or $p/$Fe event ratios, as exhibited in Figure \ref{kascade1}.
%\vskip 2mm

If protons interact with the relic background neutrinos, other cosmic ray particles may as well.   The
equivalent reactions in which $\alpha$ particles are dissociated in collisions with relic neutrinos and
antineutrinos  
\begin{eqnarray}
~~~~~~~~~~~~~~~~~~~~~~~~~~~~~~~~~~~~\alpha + \nu_e~ \rightarrow~ 3p + n + e^-\\
\alpha + \bar{\nu}_e~ \rightarrow~ p + 3n + e^+
\end{eqnarray}
have $Q$-values of 27.5 MeV and 30.1 MeV, respectively. The threshold
energies for these reactions are larger than the threshold energy
for reaction (\ref{ibeta}) by factors of 60.7 and 66.4, respectively.

If we now replace the imagined proton beam in our Gedanken experiment by a beam of $\alpha$ particles and the $\bar{\nu}_e$
target by one that consists of a mixture of $\nu_e$ and $\bar{\nu}_e$, we may expect to see the following when the beam energy 
is increased. As the energy exceeds the thresholds for the mentioned reactions, $\alpha$ particles will start to disappear from
the beam. They are replaced by protons and neutrons. The neutrons decay after a while into protons, so that each $\alpha$
particle turns into 4 protons, each of which carries, on average, 1/4 of the energy of the $\alpha$ particle. As the beam energy
increases, an increasing fraction of the $\alpha$s will undergo this process and the beam is thus increasingly enriched with
protons.
\begin{figure}[htb]
\epsfysize=8.64cm
\centerline{\epsffile{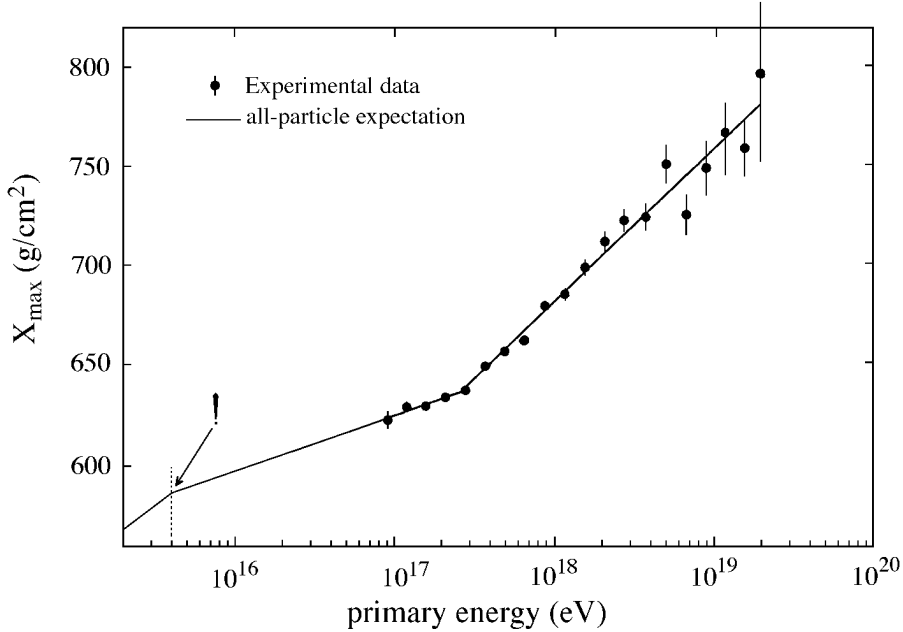}}
\caption{\small
The expected average depth of the shower maximum in the atmosphere ($a$) and the expected spectral index
($b$), as a function of the energy of the cosmic rays, predicted on the basis of extrapolations of experimental
data obtained with the Fly's Eye detector \protect\cite{Wig00}. See text for details.}
\label{fly2}
\end{figure}

Also this scenario is in detailed agreement with the experimental cosmic ray data at energies above $10^{17}$ eV.
At 100 PeV, the cosmic ray spectrum is dominated by $\alpha$ particles, since the protons have fallen victim to reaction
(\ref{ibeta}). However, as the threshold near 300 PeV is crossed, $\alpha$s start to disappear and are increasingly replaced by
protons.

We would like to point out that this explanation of the cosmic ray data in the 0.1 -- 1000 PeV energy range was already proposed
at a conference in 1999 \cite{Wig00}. At that time, neither the CASA-BLANCA (Figure \ref{xmax}) nor the KASCADE
(Figure \ref{kascade1})  results were in the public domain. Based on the data available at that time, the kink in the 
$\langle X_{\rm max}\rangle$ distribution near 4 PeV was explicitly predicted, as illustrated in Figure \ref{fly2}. 
%\vskip 2mm

The precision of the {\em neutrino mass} value that can be derived from these data is directly determined by the precision with
which the energy of the knee is known. The value of $E_{\rm knee} = 3 \pm 1$ PeV, which we adopted on the basis of the different
reported values (see Section 2.1), translates on the basis of Equation \ref{com3} into the following value for the $\nu_e$ mass:
$m_{\nu_e} = 0.5 \pm 0.2$ eV/$c^2$. This value falls nicely within the narrowing window that is still allowed by explicit
measurements of this mass. It also falls within the window (0.1 - 1 eV/$c^2$) implicated by models that explain the
Super-GZK events through a process in which extremely energetic neutrinos of extragalactic origin interact with the relic
neutrinos in our galaxy and produce $Z^0$s \cite{pas}.

The {\em energy gap} between the thresholds for the $p\bar{\nu}_e$ and  $\alpha \nu_e,\bar{\nu}_e$ reactions is {\em
independent of the neutrino mass}. It is only determined by the $Q$-values of the various reactions: $\Delta \log{E} = 1.78 ,
1.82$, in excellent agreement with the measured energy gap between the two kinks in the $\langle X_{\rm max}\rangle$ distribution
($\Delta
\log{E} = 1.8$, see Figure \ref{xmax}). This is perhaps the most remarkable and strongest point in favor of the described
scenario.

\section{A possible scenario for PeV cosmic ray production}

We now turn our attention to an extremely crucial question: {\em How could the process that forms the basis
of our hypothesis (inverse $\beta$-decay) play such a significant role, given its extremely small cross section?} 

The cross section for  $\bar{\nu}_e$ scattering off protons was measured for energies just above the
threshold ($Q = 1.805$ MeV) to be \cite{Perkins}:
\begin{equation}
\sigma~{(\bar{\nu}_e + p \rightarrow n + e^+)}~\simeq~ 10^{-43} E^2~~ {\rm cm}^2
\label{xsec}
\end{equation}
where $E$ is the $\bar{\nu}_e$ energy, expressed in units of MeV. If $m_{\nu_e} \approx 0.5$ eV, the cross section 
for the process $p + \bar{\nu}_e \rightarrow n + e^+$ is expected to scale with $E_p^2$ for
protons in the energy range between $10^{16}$ eV and $10^{17}$ eV, where the effects of this process on the energy spectra and
the elemental composition supposedly play an important role \cite{Beacom}. For a target density of $\sim 100~\nu_e$~cm$^{-3}$,
the expected cross sections ($10^{-42} - 10^{-40}$ cm$^2$) translate into mean free paths of $10^{38 - 40}$ cm, or average
lifetimes of
$10^{20 - 22}$ years, \ie\ 10 -- 12 orders of magnitude longer than the age of the Universe.
If this were all there is, the high-energy cosmic ion spectra could thus never have been affected at a
significant level by the hypothesized processes.
%\vskip 2mm

However, it is important to realize that, with a mass of 0.5 eV, the relic
$\bar{\nu}_e$s would be {\em nonrelativistic} ($kT \sim 10^{-4}$ eV). 
Typical velocities would be $< 100$ km/s in that case \cite{TG}, less than the 
escape velocity from the surface of the Sun.
Such neutrinos may be expected to have accumulated in gravitational potential wells.
Weiler \cite{weiler} has estimated that the density of relic neutrinos in our own galaxy
would increase by four orders of magnitude (compared to the universal density of 100 cm$^{-3}$)
if their mass was 1 eV.

Locally, this effect could be much more spectacular. Extremely dense objects, such as neutron
stars or black holes, could accumulate very large densities of relic neutrinos
and antineutrinos in their gravitational fields. 
Let us consider, as an example, a typical neutron star, with a mass ($M$) of $3 \cdot 10^{30}$
kg and a radius of 10 km. Even at a distance ($r$) of one million kilometers from this object,
the escape velocity is still considerably larger than the typical velocity of these relic
neutrinos: 700 km/s.
\begin{figure}[htb]
\epsfysize=7.37cm
\centerline{\epsffile{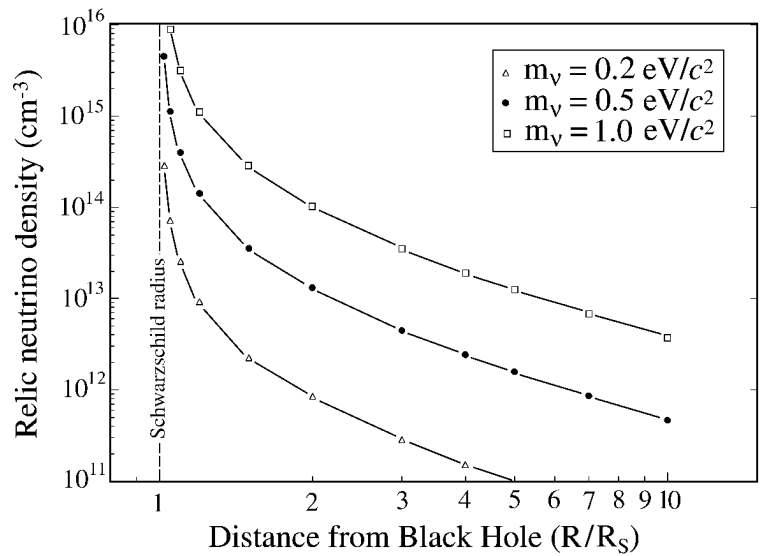}}
\caption{\small
The maximum density of neutrinos in the vicinity of a black hole, for different values of the neutrino mass.
In calculating this density, all quantum states up to the Fermi level, determined by the local escape velocity, were assumed to
be filled.}
\label{nudens}
\end{figure}

The concentration of relic neutrinos in such a local potential well is governed by
the Pauli principle, which limits their phase-space density to $4 g_\nu h^{-3}$ \cite{TG},
where $g_\nu$ denotes the number of helicity states and $h$ Planck's constant (see also Equation \ref{nuspec}). 
Since the escape velocity scales with $r^{-1/2}$, the maximum neutrino density,
$$\rho_{\nu} ({\rm max}) = \int_0^{p_{\rm esc}} N(p) dp \sim p_{\rm esc}^3$$ 
is proportional to $r^{-3/2}$, and reaches values of the order of $10^{12}~ {\rm cm}^{-3}$ near the surface of this
neutron star. If the source of the potential well has a different mass, the
achievable neutrino density scales with $M^{3/2}$.  
In the ``neutrino atmosphere'' surrounding a massive black hole, the density may become as high as 
$10^{14}~{\rm cm}^{-3}$ near the Schwarzschild radius (see Figure \ref{nudens}). The average lifetime of a 10 PeV proton
traveling in such an atmosphere would be of the order of $10^9$ years, and correspondingly shorter for even higher energies
(Equation \ref{xsec}).

This means that the accelerated cosmic protons would have to spend a very long time in this dense neutrino atmosphere 
in order to make the reaction $p + \bar{\nu}_e \rightarrow n + e^+$ a significant process. This would only be possible if the 
degenerate object in the center of this neutrino atmosphere were at the same time also the source of these accelerated particles.
This might very well be the case \cite{Hillas}. Neutron stars
usually rotate very fast and exhibit very strong magnetic fields (up to $\sim 10^8$ T).
When the magnetic axis does not correspond to the rotation axis, the changing magnetic
fields in the space surrounding the neutron star may give rise to substantial electric fields,
in which charged particles may be accelerated to very high energies. The synchrotron radiation emitted by
accelerated electrons which constitutes the characteristic pulsar signature of these objects bears witness to this phenomenon.
As an example, we mention the Crab pulsar, which is believed to be capable of accelerating protons to energies of 50 PeV and Fe
ions to 1000 PeV \cite{Hillas}.

So here follows our hypothesized scenario for the ``Great Cosmic Accelerator''.
\begin{itemize}
\item
During the gravitational collapse that led to the formation of a massive black hole somewhere in the center of our galaxy,
large numbers of relic neutrinos were trapped in the gravitational field of this object. As in other processes
that take place in the Universe, for example the Hubble expansion, all quantum states up to the Fermi level were filled and thus
densities of the order of $10^{14-15}~{\rm cm}^{-3}$ were reached near the Schwarzschild radius, $R_S$.
\vskip 2mm
\item
Of course, also large numbers of protons and other ions present in the interstellar gas were gravitationally trapped in this
event. However, these particles were subject to acceleration/deceleration in the very strong electromagnetic fields surrounding
the newly formed black hole. In addition, they interacted with each other through the strong force. In the (long) time that
has passed since the formation of the black hole, almost all these nuclei have either crashed into the black hole or escaped
from its gravitational field.
\vskip 2mm
\item
The only ions that did not undergo this fate are to be found in the equatorial plane, where they may be kept in closed orbits by
the Lorentz force, since the magnetic field is perpendicular to this plane. This accretion disk of accelerated ions is the
source of the PeV cosmic rays observed on Earth.
\vskip 2mm
\item
The magnetically trapped ions could escape from their orbits in one of two ways:
\vskip 1mm
\begin{description}
\item[A)] Collisions with nuclei from the interstellar gas in the vicinity of the black hole. 
The cross section for this process is approximately energy independent.
\item[B)] Collisions with (anti-)neutrinos. The cross section for this process increases with the ion's energy
(Equation \ref{xsec}).
\end{description}
\vskip 2mm
\item The rates for these two processes are determined by the product of the cross section and the target density.
Whereas the cross section of process A ($\sim 100$ mb) is 16 orders of magnitude larger than that for process B ($10^{-41}$
cm$^2$), the density of the relic neutrinos ($10^{14}$ cm$^{-3}$) may well exceed the density of interstellar gas in the
vicinity of the black hole by 16 {\em or more} orders of 
magnitude\footnote{Note that the relic neutrinos are 9 orders of magnitude more abundant than protons in the Universe.
This requirement is thus equivalent to an increase of the $\nu/p$ ratio by 7 orders of magnitude as a result of gravitational
trapping.}. 
This would be the case if the latter density were $< 10^4$ atoms per cubic meter.
In that case, the probabilities for the two processes are compatible and, therefore, they are in competition with each
other. 
\vskip 2mm
\item Above the knee (3 PeV), the source is selectively depleted of protons, because of process B. Since the cross
section for this process (and thus its relative importance, compared with process A) increases with energy, 
and since the more energetic particles are found in a region with higher $\nu$ density (Figure \ref{nudens}), the spectral
index
$n$ of the all-particle spectrum changes abruptly, from 2.7 to 3.0.
\vskip 2mm
\item Above the second knee, the source is {\em in addition} selectively depleted of $\alpha$s, and the slope parameter
increases further, from 3.0 to 3.3.            
\end{itemize}

In this scenario, the magnetically trapped ions would have to orbit the black hole for a long period of time before escaping,
typically $\sim 10^9$ years. One may wonder how that could be possible, since the effects of synchrotron radiation,
which are certainly non-negligible for these high-energy protons, might destabilize the particle orbits. In order to calculate
these effects, we need to know the radial dependence of the magnetic field strength, $B(r)$. In the following, we will assume
that $B(r) = B_0 r^{-3}$, as for the dipole fields generated by rotating neutron stars. 
Charged particles with momentum $p$ and charge $Z$ are then kept in a circular orbit by the Lorentz force if 
\begin{equation}
pr^2 = B_0 Z
\label{loreq}
\end{equation} 
Therefore, a loss of momentum, by synchrotron radiation, would increase the radius of the particle's orbit, but
would otherwise not distort the stability of the system. At the same time, such an increase would change the magnetic flux
through the current loop represented by the orbiting particle and the resulting emf would re-accelerate the particle such as to
prevent the change in its orbit (Lenz's law). 

The same feedback principle is applied in high-energy electron accelerators where synchrotron radiation losses are an important
factor. For example, the LEP $e^+e^-$ storage ring at CERN operated during its last year at energies in excess of 100 GeV. At
that energy, the (average) synchrotron radiation loss amounted to 2.8 GeV per orbit. On average, this
energy loss was compensated for by means of RF power. However, fluctuations about this average, which between two consecutive RF
cavities were of the same order as the average energy loss itself, would rapidly lead to an increase in the transverse emittance
of the beam, in the absence of a feedback mechanism. Yet, the LEP beam could easily be kept stable for a period of 24 hours.
During this period, which corresponds to $\sim 10^9$ particle orbits, (fluctuations in) the accumulated synchrotron radiation
losses amounted to $\sim 10^7$ times the particle's nominal energy.

Let us now consider, as an example, a black hole with a mass of $10^6 M_\odot$ ($R_S = 3\cdot 10^9$ m). Let us assume that 10 PeV
protons orbit this object at a distance of $10 R_S$. A magnetic field with a strength of 1 mT would be needed to provide the
centripetal force in that case. The protons would, on average, lose 2 GeV per orbit to synchrotron radiation, an orbit which
they complete in about 10 minutes. It would take such protons thus a period of $\sim 10^9$ years to accumulate a total synchroton
radiation loss equal to $10^7$ times their own energy. Taking the LEP example as guidance, we conclude that such
losses would not preclude orbit stability.

As the proton energy is further increased, the synchrotron radiation losses grow rapidly. In the above example, 100 PeV protons
orbit the black hole at a distance of $\sqrt{10} R_S$, where the magnetic field strength is 32 mT. Since the specific energy loss
$dE/dx$ scales with $E^4 r^{-2}$, these protons lose energy to synchrotron radiation at a rate that is $10^5$ times larger
than that for the 10 PeV ones. Therefore, it takes them only $10^5$ years to accumulate a total loss equivalent to $10^7$ times
their own energy. And although it might well be possible that their orbits are stable against the effects of synchrotron
radiation for a much longer period of time, we cannot derive support for that from the LEP example, as we did for the 10 PeV
protons. If the feedback mechanism were not capable to compensate completely for the synchrotron radiation losses, the
particle would gradually spiral outward and end up in an orbit where it is (sufficiently) stable against any further energy 
losses.

Because of the mentioned $dE/dx$ scaling characteristics of synchrotron radiation, it requires much less imagination to make
the described scenario work for a supermassive black hole than for a black hole that resulted from the collapse of a massive
star, say with a mass of $10 M_\odot$ ($R_S = 3\cdot 10^4$ m). In the latter case, the specific energy losses due to synchrotron
radiation would be 10 orders of magnitude larger than in the previous examples. Thus, a 100 PeV proton orbiting such a
black hole at a radius of
$5 R_S$ would lose energy at the prodigious rate of 40 TeV/m. It is unclear how and not very likely that in this case a stable
configuration could be achieved that involves protons of such high energies.      

One important aspect that we have not yet discussed is the power-law character of the energy spectra of the cosmic ray particles.
Although the described scenario does not {\sl guarantee} this characteristic feature of the experimental data, it can be shown
that a reasonable choice of the boundary conditions does lead to a power-law spectrum with approximately the right spectral
index. 
Equation \ref{loreq} shows that if $B$ behaves as a dipole field, the region between the radii $10 R$ and $R$ ($R >
R_S$) could accommodate  (ultrarelativistic) protons with energies between $E_0$ and $100 E_0$, as well as heavier nuclei with
energies between $Z E_0$ and $100~ZE_0$. The most energetic particles would be found closest to the black hole.
A constant density of accelerated particles throughout the accretion disk would then imply that $dN/dE \sim E^{-2.0}$. The
effects of synchrotron radiation and aging of the black hole would lead to a further steepening of the spectrum, \ie a further
increase of the spectral index $n$. The first effect increases the particle density at lower energies (larger radii) at the
expense of that at higher energies. The second effect is a consequence of the gradual increase in the total cross section
observed in high-energy
$pp$ collisions
\cite{Castaldi}. As a result, the source spectrum is more depleted at higher energies (smaller radii), to an extent determined by
the age of the black hole.

We would also like to point out that several pulsars are known to produce relativistic electrons with spectra that
follow a power-law. These electrons are accelerated in the same em fields that form the basis of our scenario for PeV cosmic ray
production. 

Obviously, this scenario is not supported by observational evidence of the quality
discussed in the previous sections. It is in fact little more than an imagined conspiracy of factors which, together, lead to
measurable effects of a process that stopped playing a role in the Universe at large at the tender age of one second. However,
it is {\em not inconceivable}, in the sense that no known physics principle is violated and no experimentallly observed fact is
ruled out. And apart from the fact that this scenario would make interactions between high-energy cosmic nuclei and relic
neutrinos a significant process that would explain many features of the cosmic ray spectra in the energy range from 0.1 -- 1000
PeV, it also has the merit that it provides an origin and an acceleration mechanism for the cosmic rays in this energy range.
This in contrast with the Supernova shockwave acceleration models, which run out of steam in the $10^{14}$ eV region and do not
offer any explanation for particles at higher energies.

\section{Conclusions} 
\vskip-5mm
The high-energy cosmic ray spectra exhibit some intriguing features that can all be explained in a
coherent manner from interactions between cosmic protons or $\alpha$ particles and relic $\bar{\nu}_e$s if
the latter have a restmass of about 0.5 eV/$c^2$: 
\begin{itemize}
\item Two ``knees'', \ie significant changes in the spectral index of the all-particle spectrum, which would correspond to the
thresholds for the $p\bar{\nu}_e$ and $\alpha \bar{\nu}_e$ reactions.
\vskip 2mm
\item These knees coincide with kinks in the $\langle X_{\rm max}\rangle$ distribution, which measures the average depth
inside the Earth's atmosphere at which the showers initiated by the cosmic rays reach their maximum intensity. 
\vskip 2mm
\item The measured energy separation between these kinks ($\Delta \log{E} = 1.8$) is exactly what one would expect on the basis
of the difference between the $Q$-values of the $p\bar{\nu}_e$ and the $\alpha \nu_e,\bar{\nu}_e$ reactions 
($\Delta \log{E} = 1.80 \pm 0.02$).
\vskip 2mm
\item The kinks in the $\langle X_{\rm max}\rangle$ distribution initiate changes in the elemental composition of the cosmic
rays that are in detailed agreement with the changes one should expect when the thresholds for the $p\bar{\nu}_e$ and $\alpha
\bar{\nu}_e$ reactions are crossed: A selective depletion of the proton component of the source spectrum at the first kink,
a selective depletion of $\alpha$ particles combined with a reintroduction of protons at the second kink.   
\end{itemize}

If collisions with relic neutrinos were indeed responsible for the described features, a large
concentration of such neutrinos would have to be present in the vicinity of the source of the high-energy cosmic baryons, in
order to explain the observed event rates. We have shown that the required conditions could be met if charged particles
accelerated and stored in the equatorial plane of a supermassive black hole in our galaxy were the source of the 0.1 --
1000 PeV cosmic rays measured here on Earth. This model could also explain the energy spectra of the hadronic cosmic rays.

If our model turned out to be correct, the PeV cosmic rays would provide the first direct measurement of a
neutrino mass: $m_{\nu_e} = 0.5 \pm 0.2$ eV/$c^2$. They would also provide evidence for a key aspect of the Big Bang model and
thus offer a unique window on the leptonic era. 

A crucial test of this model will be provided by the next generation of $^3$H decay experiments. The proposed KATRIN experiment
is designed to be able to measure a non-zero $\bar{\nu}_e$ mass down to values as small as $\approx 0.2$ eV/$c^2$ \cite{Mainz}
and should thus be in a position to either confirm or to rule out the mass value implied by our explanation of the
experimental features of the PeV cosmic rays.    

\bibliographystyle{unsrt}

\end{document}